\documentstyle[12pt]{article}
\evensidemargin \oddsidemargin
\def\beq{\begin{equation}}
\def\eeq{\end{equation}}
\def\bea{\begin{eqnarray}}
\def\eea{\end{eqnarray}}
\def\ba{\begin{array}}
\def\ea{\end{array}}
\def\del{\partial}
\def\b{\bar}
\def\t{\widetilde}
\def\d{\dagger}
\def\s{\sigma}

\def\al{\alpha}

\begin{document}
\pagestyle{empty}
\begin{flushright}
hep-th/9712213\\
CERN-TH/97-344\\
ICTP-IC/98/15\\
\end{flushright}
\begin{center}
\vspace{.5cm}
{\Large Gauge Theory Description of D-brane Black Holes:}\\
\vspace{.2cm}
{\Large Emergence of the Effective SCFT and Hawking Radiation}\\
\vspace*{1cm}
{\bf S. F. Hassan}{\footnote{\tt e-mail: fawad@ictp.trieste.it}} \\
\vspace{.2cm}
High Energy Physics Group, ASICTP, 34100 Trieste, Italy\\
\vspace{.5cm}
{\bf Spenta R. Wadia}{\footnote{\tt e-mail: wadia@nxth04.cern.ch, 
wadia@theory.tifr.res.in}}'{\footnote{On leave from the Tata  
Institute of Fundamental Research, Homi Bhabha Road, Mumbai-400005,
India}}'{\footnote{Also at Jawaharlal Nehru Centre for Advanced
Scientific Research, Jakkur P.O., Bangalore 560064, India}}\\
\vspace{.2cm}
Theoretical Physics Division, CERN,
CH - 1211 Geneva 23, Switzerland \\
\vspace*{1cm}
{\bf Abstract}
\begin{quote}
We study the hypermultiplet moduli space of an N=4, $U(Q_1)\times
U(Q_5)$ gauge theory in 1+1 dimensions to extract the effective SCFT
description of near extremal 5-dimensional black holes modelled by a
collection of D1- and D5-branes. On the moduli space, excitations with
fractional momenta arise due to a residual discrete gauge invariance.
It is argued that, in the infra-red, the lowest energy excitations are
described by an effective $c=6$, N=4 SCFT on $T^4$, also valid in the
large black hole regime. The ``effective string tension'' is obtained
using T-duality covariance. While at the microscopic level, minimal
scalars do not couple to (1,5) strings, in the effective theory a
coupling is induced by (1,1) and (5,5) strings, leading to Hawking
radiation. These considerations imply that, at least for such black
holes, the calculation of the Hawking decay rate for minimal scalars 
has a sound foundation in string theory and statistical mechanics and,
hence, there is no information loss.
\end{quote}
\end{center}
\vfill
\begin{flushleft}
hep-th/9712213\\
CERN-TH/97-344\\
ICTP-IC/98/15\\
December 1997\\
\end{flushleft}
\vfill\eject
\pagestyle{plain}
\section{Introduction and Summary}

In the recent past there has been encouraging progress in
understanding the statistical basis of black hole entropy
\cite{SUSS,SEN,StVa}. This progress is in part due to new developments
in superstring theory, both technical and conceptual. The technical
part has to do with supersymmetry, which brings some non-perturbative
aspects under control, and the conceptual part has to do with D-branes
and duality symmetries of string theory. For a review see
\cite{Pol,SCHWARZ}. The fact that there is a statistical basis for the
black hole entropy means that one has understood what is usually
dubbed as {\it intrinsic gravitational entropy} \cite{HAWPEN}, in
terms of degrees of freedom hitherto unknown within general
relativity. As yet we do not know how these new degrees of freedom are
inscribed in the metric and other long range fields and hence, as yet,
we do not ``understand'' the geometric Bekenstein-Hawking entropy
formula. However, even with this lacking, we can attempt to answer
some of the conceptual issues raised by black hole thermodynamics. The
conceptual issue is the so called {\it information paradox} which says
that black hole radiation is exactly thermal \cite{HAWPEN}. Such an
assertion, made within the standard formulation of general relativity,
denies a statistical basis of black hole thermodynamics. A statistical
basis explains thermodynamics in terms of the statistical averages of
unitary amplitudes, in which case, information loss is not intrinsic.
Such a view has been advocated by 't Hooft \cite{THOOFT}. The success
of the statistical derivation of black hole entropy has suggested a
derivation of black hole thermodynamics in terms of some constituent
degrees of freedom of the black hole
\cite{CaMa,MaSu,DMW,DM1,MaSt,CGKS,KK,MALDA,MaTh,JKM,ANG}. In most of
these studies, for technical reasons, one focuses on the near extremal
black hole of type IIB string theory compactified on a 5-torus in a
particular range of parameters. In this range of parameters the black
hole has positive specific heat. Even in this specialised situation,
the black hole has a large number of degrees of freedom and a study of
black hole thermodynamics leads to the ``information paradox'' as the
black hole radiates to its zero temperature ground state. A complete
solution of even this simplified problem is not easy and up to now all
calculations of black hole thermodynamics have been performed when the
effective open string coupling is small, and one is not in the large
black hole regime. However the precise agreements of the grey body
factors at long wave lengths encourage us to search for a precise
microscopic description of the black hole which is valid even when the
effective open string coupling is large.

This paper is devoted to a study of the D-brane model of the
5-dimensional black hole of IIB string theory with charges $Q_1$,
$Q_5$ and $N$ \cite{CaMa}. The model consists of $Q_1$ D1-branes and
$Q_5$ D5-branes wrapped around $S^1\times T^4$, carrying excitations
of total momentum $N$ along $S^1$ of radius $R$. At low energies, this
model is described by a $U(Q_1)\times U(Q_5)$ N=(4,4) super Yang-Mills
theory on a 2-dimensional cylinder of radius $R$. When the black hole
is macroscopic ($gQ_1, gQ_5>>1$), this gauge theory is strongly
coupled. A counting argument in \cite{CaMa,MaTh} indicates that the
microscopic degrees of freedom of the black hole correspond to
hypermultiplets originating in the (1,5) string sector of the D-brane
system. For the D-brane configuration to appear as a black hole in the
four directions transverse to $S^1\times T^4$, $R$ must be much
smaller than the radius of the extremal black hole (the $T^4$ radii
are taken to be of string size). On the other hand, as noted in
\cite{MaSu}, in order to explain the black hole thermodynamics, the
D-brane system must have excitations of energy much lower than
$1/R$. To resolve this problem, the picture of ``multiply wound''
branes was suggested in \cite{MaSu}, based on an observation in
\cite{DM0}. This amounts to replacing the radius $R$ by $RQ_1Q_5$. If
the black hole entropy is related to the degeneracy of states in a
superconformal field theory then, in order to get the right entropy,
the central charge must be set to $c=6$. One may try to implement the
notion of ``multiple winding'' of D-branes by introducing a Wilson
line in the Weyl group of the corresponding gauge theory. This,
however, does not seem to lead to a consistent description. An earlier
attempt \cite{FHSW} to explain the black hole degrees of freedom in
terms of the low-energy excitations of the gauge theory involved a
variant of this approach, with Wilson lines in the centre of the gauge
group. However, as will be shown here, such constructions are
unnecessary and the fractionalization of momentum is a consequence of
a residual gauge invariance in the theory which leads to the existence
of sectors with twisted-periodic boundary conditions on $S^1$.

Another approach followed to investigate this D-brane system and its
coupling to bulk fields is based on a variant of the Dirac-Born-Infeld
action for a D-string on $S^1$ \cite{DM1,CGKS}. While a DBI action for
the D1,D5-brane system is not known, one starts with a DBI action for
a single D-string with 4 out of its 8 transverse oscillations frozen
to simulate the effect of a single D5-brane. To include the effect of
multiple D-branes, one resorts to the ``multiple winding'' picture and
enlarges the radius of the circle $S^1$ from $R$ to $RQ_1Q_5$.
In this way, the black hole is modelled by an effective D-string.
Surprisingly, this heuristic construction leads to a rather successful
model of the black hole in which many calculations have been
performed, although the microscopic origins of this model are not very
clear. On expanding the DBI action constructed in this way, one
obtains  
\beq
S_{DBI}= T_{eff}\int dt\int_0^{2\pi RQ_1Q_5} d\sigma
\del_\al X^{m}\del^\al X^{m}\qquad +\quad {\mbox{couplings}}\,.
\label{Sdbi}
\eeq
This, to lowest order, and after including fermions, is a $c=6$
superconformal field theory with its target space as the $T^4$
transverse to the D1-brane. The fields $X^m$ are the transverse
oscilations of the D-string in this $T^4$ which, in the microscopic
picture, would be related to the (1,1) strings. This is at variance
with the expectation that the black hole degrees of freedom have their
origin in the (1,5) string sector of the D-brane system. Furthermore,
the effective string tension $T_{eff}$ cannot be derived in this
framework. Also, the form of the couplings to bulk fields obtained in
this way do not lead to correct results for fixed scalars \cite{KK},
though this is not the case with coupling to minimal scalars.

In this paper, we study the $U(Q_1)\times U(Q_5)$ gauge theory for the
D-brane system in a systematic way and isolate the effective theory
for its low-energy excitations that are relevant to the low-energy
dynamics of the near extremal black hole in D=5.  The content of the
paper is organised as follows: In section 2, we describe the relevant
features of the hypermultiplet sector of $U(Q_1)\times U(Q_5)$ gauge
theory which describes the low-energy dynamics of the D1,D5-brane
system. The coupling of the (1,5) hypermultiplets and, partly, the
field content of the theory is fixed by imposing covariance under a
set of T-dualities that interchanges the D1- and D5-branes. 

In section 3, we describe a parametrization of the space of solutions
${\cal M}_0$ to the conditions for the vanishing of the D-term
potential. After gauge fixing, we show that, generically, this space
can be almost entirely parametrized in terms of the
(1,5) hypermultiplets, while the (1,1)  and (5,5) hypermultiplets
induce a metric on it. After gauge fixing, we are still left with a
discrete residual gauge group $S(Q_1-1)\times S(Q_5-1)$ which is a
subgroup of the Weyl group of $U(Q_1)\times U(Q_5)$, and maps ${\cal
M}_0$ to itself. The hypermultiplet moduli space is not renormalized
and therefore this description is valid even in the strong coupling
limit of the gauge theory which is the regime of macroscopic black
hole. 

In section 4, we consider oscillations of the moduli fields for which
the D-term potential stays zero. These are the massless excitations of
the theory on the Higgs branch and are relevant to the black hole
degrees of freedom. The residual discrete gauge invariance $S(Q_1-1)
\times S(Q_5-1)$ enables us to impose twisted-periodicity conditions
on the moduli fields on $S^1$. As a result, many components of the
moduli fields are sewn into one single field which is periodic on a
circle of larger radius, and hence has fractional momentum on the
original space. This is how very long wavelength excitations
emerge. Our sewing procedure generalizes the one used in
\cite{DVV2,DVV1,DVV3,MOTL} for mutually commuting square matirces, to
arbitrary rectangular matrices. The non-linear sigma-model on the
moduli space, when written in terms of sewn variables, takes a very
complicated and non-local form that cannot be analysed directly. We
are interested in the infra-red limit of this model, obtained after
integrating out all higher momentum modes and retaining only the
lowest ones. Assuming that the infra-red theory is local, N=4
supersymmetry along with the compactness of the moduli space and some
general considerations leads us to a $c=6$ superconformal field theory
on a target space $T^4$. This $T^4$ is different from the one in
(\ref{Sdbi}) and is not part of the 10-dimensional space-time. The
basic variable is the renormalized form of the sewn (1,5) field with
the lowest momentum quantum $\sim 1/RQ_1Q_5$. We regard this field as
an order parameter for low-energy excitations of the system in the
infra-red. The SCFT has an $SO(4)$ symmetry, instead of an $SU(2)_R$
of the gauge theory. This SCFT is also valid in the strong coupling
regime of the gauge theory $gQ_{1,5}>1$.

In section 5, we discuss the connection between this $c=6$ SCFT and
the black hole in some more detail: The ``effective string tension'',
$T_{eff}$, that has so far eluded a consistent derivation, is related
to the coupling of (1,5) hypermultiplets that is fixed by T-duality
and is given by $T_{\it {eff}}= \frac{1}{\al'^2}\sqrt{\frac{V_4} {g^2
Q_1Q_5}}$. This is different from what one would expect if the
``effective string'' is interpreted as a D-string along the $x^5$
direction. But it is consistent with the ``mean string'' picture
suggested in \cite{MATHUR}, based on the requirement that the
effective string produces (at least in principle) the correct cross
section for higher angular momentum scattering from the black hole.
We then describe the identification of the extremal and near extremal
black holes in terms of states in the SCFT with degeneracies related
to the entropy. For a given extremal black hole, the SCFT states that
do not have a near-extremal black hole interpretation are
automatically removed by a level matching condition. This condition
originates in the residual discrete gauge invariance of the theory on
the moduli space and insures the consistency of the description.
Coupling to bulk fields are then given by SCFT operators allowed by
the level matching condition.

Next, to make contact with Hawking radiation, we discuss the coupling
of the SCFT to minimal scalars in the bulk. A generic coupling can be
written using the $SO(4)$ invariance of the SCFT emerging in the
infra-red limit. However, while the black hole degrees of freedom are
contained in the (1,5) hypermultiplets, these, as shown in \cite{Ha},
do not couple to minimal scalars at the microscopic level. In our
approach, we can easily see that an effective coupling of the minimal
scalars to the black hole degrees of freedom is induced through the
coupling of these scalars to (1,1) and (5,5) hypermultiplets. The
calculation of emission and absorption rates for these scalars is, in
its technical aspects, the same as before \cite{CaMa,DMW,DM1,MaSt} and
hence one can exactly reproduce the grey body factors as calculated in
the semi-classical approach to general relativity. We also comment on
the range of validity of comparisons between the D-brane and the
thermodynamic descriptions of this black hole. As the black hole
radiates and approaches the extermal limit, the thermodynamic
description breaks down \cite{Pr} (since temperature fluctuations blow
up) while the SCFT description based on the D-brane model is still
valid. Section 6 contains the conclusions.

\section{The SUSY Gauge Theory for the D-brane Model of the Black
Hole} 

Type IIB string theory with five coordinates, say $x^5\cdots x^9$,
compactified on $S^1\times T^4$, admits a black hole solution in the
five non-compact directions $x^0=t,x^1,\cdots,x^4$. This 5-dimensional
black hole carries RR charges $Q_1$ and $Q_5$, and a Kaluza-Klein
charge $N$ corresponding to a momentum along the $S^1$. In the
extremal limit (and in the near extremal region) it is modelled by a
collection of low-energy states in a system of $Q_1$ D1-branes and
$Q_5$ D5-branes \cite{CaMa,MaTh}. The D1-branes are parallel
to the $x^5$ coordinate compactified to a circle $S^1$ of radius $R$,
while the D5-branes are parallel to $x^5$ and $x^6,\cdots,x^9$
compactified on a torus $T^4$ of volume $V_4$. The charge $N$ is
related to the momenta of very low-energy excitations of this system
along $S^1$. We take the $T^4$ radii to be of the order of $\alpha'$
and smaller than $R$ which, in turn, is much smaller than the black
hole radius. The low-energy dynamics of this D-brane system is
described by a $U(Q_1)\times U(Q_5)$ gauge theory in two dimensions
with $N=4$ supersymmetry. In this section, we will describe some
aspects of this gauge theory that are relevant to the identification
of the black hole degrees of freedom in the D-brane system.

The elementary excitations of the D-brane system correspond to open
strings with two ends attached to the branes and there are three
classes of such strings: the (1,1), (5,5) (1,5) strings. The
associated fields fall into vector multiplets and hypermultiplets,
using the terminology of N=2, D=4 supersymmetry. In the following, we
will only consider the hypermultiplet sector since this sector
contains the low-energy black hole degrees of freedom we are
interested in. The part of the spectrum coming from (1,1) strings is
simply the dimensional reduction, to $1+1$ dimensions (the
$(t,x^5=\s)$-space), of the N=1 $U(Q_1)$ gauge theory in $9+1$
dimensions \cite{Pol,Wit1}.  The gauge field components
$X^{(1)}_m(\s,t)$ ($m=6,7,8,9$) along the $T^4$, together with their
fermionic superpartners, form a hypermultiplet in the adjoint of
$U(Q_1)$, while the remaining components form a vector
multiplet. Since $x^m$ are compact, the (1,1) strings can also have
winding modes around the $T^4$. These are, however, massive states in
the $(1+1)$-dimensional theory and can be ignored. Similarly, the part
of the spectrum coming from (5,5) strings is the dimensional
reduction, to $5+1$ dimensions, of the N=1 $U(Q_5)$ gauge theory in
$9+1$ dimensions. In this case, the gauge field components $A^{(5)}_m$
($m=6,7,8,9$) also have a dependence on $x^m$. A set of four T-duality
transformations along $x^m$ interchanges D1- and D5-branes and also
converts the momentum modes of the (5,5) strings along $T^4$ into
winding modes of (1,1) strings around the dual torus \cite{WT}. Since
these winding modes have been ignored, a T-duality covariant
formulation requires that we should also ignore the associated
momentum modes. Thus we can only retain the zero modes of $A^{(5)}_m$
along $T^4$, denoted by $X^{(5)}_m(\s,t)$. These fields fall in a
hypermultiplet in the adjoint of $U(Q_5)$, while the remaining fields
form a vector multiplet.

The field content obtained so far is that of N=2 $U(Q_1)\times U(Q_5)$
gauge theory in 1+5 dimensions, reduced to 1+1 dimensions on $T^4$.
The $SO(4)\sim SU(2)_L\times SU(2)_R$ rotations of the torus act on
the components of the adjoint hypermultiplets $X^{(1,5)}_m$ as an
$R$-symmetry. To this set of fields we have to add the fields from the
(1,5) sector that are constrained to live in 1+1 dimensions by the ND
boundary conditions. These strings have their ends fixed on different
types of D-branes and, therefore, the corresponding fields transform
in the fundamental representation of both $U(Q_1)$ and $U(Q_5)$. The
(1,5) sector fields also form a hypermultiplet but with only $SU(2)_R$
as the $R$-symmetry group. We denote these fields by $\chi_i(\s,t)$,
where $i$ is the $SU(2)_R$ doublet index. The inclusion of these
fields breaks the supersymmetry by half, to the equivalent of N=1 in
D=6, and the final theory only has an $SU(2)_R$ $R$-symmetry. To make
this manifest, we write the hypermultiplets $X^{(1,5)}_m$ in terms of
$SU(2)_R$ doublets $N^{(1,5)}_i$ given by (see for example, \cite{Na}) 
\beq 
\sigma^m X_m =
\left(\ba{rl} X_9 + iX_8 & X_7 + iX_6 \\ -X_7 +iX_6 & X_9 - iX_8
\ea\right) =\left(\ba{rl} N_1 & N_2\\ -N_2^\d & N_1^\d \ea\right)\,.
\label{XN}
\eeq
Here, $\sigma^m = (i\tau^1,i\tau^2,i\tau^3,{\bf 1})$, and $\tau^I$
are the Pauli matrices. 

At low-energies, a lagrangian for the hypermultiplets $\chi_i$,
$N^{(1)}_i$ and $N^{(5)}_i$ as well as the two vector multiplets can
be easily written by constructing an N=2 $U(Q_1)\times U(Q_5)$ gauge
theory in $3+1$ dimensions and then reducing it to $1+1$ dimensions
(see for example, \cite{West}). However, note that $N^{(1,5)}_i$ are
related to the zero modes of gauge fields an the compact space $T^4$ either
directly, or by T-duality. As a result, these fields are valued on a
compact space. This information is not contained in the field theory
and we will impose it as an extra condition on the field variables. In
the following we will only write down some relevant terms of this
lagrangian which are needed for the analysis in the next sections. But
first, some notation: The fundamental representation indices for the
gauge group $U(Q_5)$ are denoted by $a,b,\dots$ and those for $U(Q_1)$
are denoted by $a',b',\dots$. For the adjoint representations, we use
the indices $s$ and $s'$, respectively. The indices $i,j$ label the
fundamental doublet of $SU(2)_R$ and its generators are denoted by
$\tau^I/2$.  Thus in components, the scalars in the hypermultiplets
take the form $\chi_{ia'a}$, $N_{ia'b'}^{(1)}$ and
$N_{iab}^{(5)}$. Under a gauge transformation, $\chi_i$ transform as
$\chi_i\rightarrow U_1\chi_i U^{-1}_5$ where, $U_5\in U(Q_5)$ and
$U_1\in U(Q_1)$. Also, $\al=0,1$ labels the coordinates on the
$t,x^5=\s$ space.

The only terms needed for the study of dynamics on the Higgs branch
of this theory are the kinetic energy terms for the hypermultiplets
and the D-term potential in the theory. The kinetic energy terms are
given by 
\beq
S_{ke}
=k_{11}{\rm Tr}\int d^2\sigma\del_\al N^{(1)}\del^\al {\b N}^{(1)} 
+k_{15}{\rm Tr}\int d^2\sigma \del_\al \chi \del^\al {\b\chi}
+k_{55}{\rm Tr}\int d^2\sigma \del_\al N^{(5)}\del^\al {\b N}^{(5)}\,,
\label{one-IV} 
\eeq
where, $(\b\chi)^i=\chi^\d_i$ etc., and the traces are normalized to
identity. All fields have been scaled using powers of $\al'$
such that they have dimensions of length. With this convention, the
couplings are given by 
\beq
k_{11}=\frac{1}{\al'gQ_1}\,,\quad
k_{15}=\frac{1}{\al'^2g}\sqrt{\frac{V_4}{Q_1Q_5}}\,,\quad
k_{55}=\frac{V_4}{\al'^3gQ_5}\,.
\label{kkk}
\eeq
where, $g$ is the string coupling constant. As we will see, $k_{15}$
is the only coupling that appears in the effective low-energy theory
describing the near extremal dynamics of the black hole and is often
referred to as the ``effective string tension''. Its value has been
fixed by a simple T-duality argument: Since the hypermultiplets
correspond to D-brane excitations, we know that $k_{11}\sim
(\al'g)^{-1}$ and $k_{55}\sim V_4(\al'^3g)^{-1}$. The $V_4$ arises
from the reduction, on $T^4$, of the 5-brane worldvolume theory and
$\al'$ takes care of the dimensions. In general, $k_{15}$ will be of 
the form $c(\al'g)^{-1}$, where $c$ can only be a function of the
dimensionless quantity $V_4\al'^{-2}$. Under T-duality along all $T^4$
directions, $V_4$ and $g$ transform as
\beq
V'_4 = \frac{\al'^4}{V_4}\,, \qquad
g'= g \frac{\al'^2}{V_4}\,.
\label{two-IV}
\eeq
This interchanges $k_{11}$ and $k_{55}$ as a consequence of the
interchange of D1-branes and D5-branes under such a duality. 
However, the (1,5) string sector remains unchanged, implying that
$c/g$ must go over to itself. This requirement fixes 
$c=\sqrt{V_4/\al'^2}$. Furthermore, the gauge theory is studied in the
limit of $g\rightarrow 0$ and $Q_1, Q_5\rightarrow \infty$ such that
$gQ_1$ and $gQ_5$ are finite. To write the theory in a meaningful way,
we scale the fields appropriately so that the action depends on the
well defined finite couplings. Taking the T-duality covariance into
account, this amounts to scaling $\chi$ by a factor of
$(Q_1Q_5)^{-1/4}$ and $N^{(1,5)}$ by $(Q_{1,5})^{-1/2}$. This leads to
the couplings as given in (\ref{kkk}).

The lagrangian also contains a D-term potential 
$k_{11}D^{(1)Is'}D^{(1)Is'}+k_{55}D^{(5)Is}D^{(5)Is}$, with the
D-terms given by  
\bea
k_{11}D^{(1)Is'}&=& \tau^{Ii}_j{\rm Tr}\left\{T^{s'}
\left(k_{15}\chi_i \b\chi^j+k_{11}[N^{(1)}_i,{\b N}^{(1)j}]\right)
\right\}\,,
\label{B-III} \\
k_{55}D^{(5)Is}&=& \tau^{Ii}_j{\rm Tr}\left\{T^{s}
\left(k_{15}\b\chi^j\chi_i+k_{55}[N^{(5)}_i,{\b N}^{(5)j}]\right)
\right\}\,.
\label{B-IV}
\eea 
Here, $T^{s'}_{a'b'}$ and $T^{s}_{ab}$ are the generators of $U(Q_1)$
and $U(Q_5)$ respectively. In the future, we will suppress the
couplings in the D-terms. The remaining terms of the lagrangian, that
we have omitted, correspond to two vector multiplets and their
couplings to the hypermultiplets and are not needed for our
analysis. We will first consider this theory in the perturbative
regime where $gQ_{1,5}<1$ and then argue that our results can safely
be extrapolated to the large black hole regime where $gQ_{1,5}>1$. As
stated above, $N^{(1,5)}$ take values on a compact space.  Note that,
in the limit $V_4\rightarrow \infty$, the (5,5) sector decouples and
the gauge group $U(Q_5)$ reduces to a flavour group. Since $T^4$ goes
over to $R^4$, the fields $N^{(1)}$ are no longer compact. This is the
theory analyzed in \cite{Wit2}.

\section{The Moduli Space}

Our aim is to study the low-lying excitations of the gauge theory
described in the previous section, in the strong coupling regime
$gQ_{1,5}>1$. This is the macroscopic black hole regime of the D-brane
system. However, to begin with, we consider the system in the regime
$gQ_{1,5}<1$ which is the perturbative regime of the gauge theory. To
isolate the massless excitations (to be later identified as the black
hole degrees of freedom) we restrict ourselves to the Higgs phase
where all fields are set to zero except for the hypermultiplets
$\chi_i$, $N^{(1)}_i$ and $N^{(5)}_i$, altogether containing $4(Q_1^2
+ Q_5^2 + Q_1Q_5)$ components.  We then look at configurations of
these hypermultiplets for which the D-terms vanish. For constant field
configurations, this would normally define the moduli space of vacua
on the Higgs branch, though in 1+1 dimensions this notion is not well
defined due to strong infra-red fluctuations. However, we are
interested in space-time dependent configurations of fields for
which the D-terms vanish, and use the term ``moduli space'' only in
this sense. After gauge fixing, these configurations correspond to the
independent low-energy degrees of freedom relevant to the black hole
problem. In this section, we describe how these degrees of freedom are
obtained by setting the D-term potential to zero and fixing the
gauge. 

The D-terms in (\ref{B-III}),(\ref{B-IV}) are of the form $D^{Is}={\rm
Tr}(T^s D^I)$, where $D^I$ are hermitian matrices. The generators
$T^s$ also include the identity, corresponding to the overall $U(1)$
factors in $U(Q_{1,5})$. Therefore, $D^{Is}=0$ implies $D^I=0$. Thus, the
vanishing of the D-terms (\ref{B-III}),(\ref{B-IV}) leads to the
following sets of equations (for $I=1,2,3$ and with the couplings
suppressed): 
\bea
&\tau^{Ii}_j\,\left(\chi_i\b\chi^j+\,[N^{(1)}_i\,,\,{\b N}^{(1)j}]
\right)_{a'b'}=0&\,, 
\label{II-four} \\
&\tau^{Ii}_j\,\left(\b\chi^j\chi_i+\,[N^{(5)}_i\,,\,{\b N}^{(5)j}]
\right)_{ab} = 0 &\,.
\label{II-five}
\eea
Equation (\ref{II-four}) has its origin in the $U(Q_1)$ sector of the
theory and, using the standard representation of Pauli matrices
$\tau^I$, it has the following independent components
\bea
&(\chi_1\chi_1^\d - \chi_2\chi_2^\d)_{a'b'}+[N^{(1)}_1\,, 
\,N^{(1)\d}_1]_{a'b'}-[N^{(1)}_2\,,\,N^{(1)\d}_2]_{a'b'}= 0\,,& 
\label{II-six} \\
& (\chi_1\chi_2^\d)_{a'b'}+[N^{(1)}_1\,,\,N^{(1)\d}_2]_{a'b'}=0\,.& 
\label{II-seven}
\eea
The first equation is real while the second one is complex, thus there
are $3Q_1^2$ constraints coming from this set of equations. Similarly,
equation (\ref{II-five}) comes from the $U(Q_5)$ sector of the gauge
theory and gives rise to the $3Q_5^2$ constraints, 
\bea
&(\chi_1^\d\chi_1 - \chi_2^\d\chi_2)_{ab}+[N^{(5)}_1\,, 
\,N^{(5)\d}_1]_{ab}-[N^{(5)}_2\,,\,N^{(5)\d}_2]_{ab} = 0\,,&
\label{II-eight} \\
&(\chi_1^\d\chi_2)_{ab}+[ N^{(5)}_2\,,\, N^{(5)\d}_1]_{ab}=0\,.& 
\label{II-nine}
\eea
Equations (\ref{II-six},\ref{II-seven}) and 
(\ref{II-eight},\ref{II-nine}) have the same trace parts corresponding
to the vanishing of $U(1)$ D-terms, namely,  
\beq
\chi_{1a'a} \chi^*_{1a'a} - \chi_{2a'a} \chi^*_{2a'a}= 0
\,,\qquad
\chi_{1a'a} \chi^*_{2a'a}=0\,,
\label{II-twelve}
\eeq
which are three real equations. Therefore, the vanishing of D-terms
imposes $3Q_1^2 + 3Q_5^2-3$ constrains on the fields. Gauge fixing can
remove another $Q_1^2+ Q_5^2-1$ scalar field components, thus leaving
only $4(Q_1Q_5+1)$ components to parametrize the moduli space. The
explicit structure of this moduli space is important for our analysis
and will be described below.

Let us start with gauge fixing by gauging away components of the
adjoint hypermultiplets $N^{(1,5)}_i$. For this, it is convenient to
parametrize these complex fields in terms of the hermitian metrices
$X^{(1,5)}_m$ as in (\ref{XN}). In this parametrization, we can use
the $U(Q_1)$ transformations to diagonalize one of the $X^{(1)}_m$,
say, $X^{(1)}_6$. This fixes $U(Q_1)$ down to $U(1)^{Q_1}$. From the
remaining $U(1)$'s, $Q_1-1$ of them can be used to gauge away that
number of phases from any other $X^{(1)}_m$, say $X^{(1)}_7$. Let us
gauge away the phases of the components $(X^{(1)}_7)_{1\t b'}$ where
$\t b'=2,\cdots, Q_1$. The remaining $U(1)$ factor is the overall
abelian factor of $U(Q_1)$ which leaves fields in the adjoint
representation invariant. Similarly, we can fix $U(Q_5)$ down to its
overall $U(1)$ subgroup by diagonalizing $X^{(5)}_6$ and removing the
phases from $(X^{(5)}_7)_{1\t b}$, where $\t b=2,\cdots, Q_5$.  After
fixing the gauge (down to a $U(1)\times U(1)$ subgroup) in this
manner, $X^{(1,5)}_8$ and $X^{(1,5)}_9$ remain arbitrary hermitian
matrices, while $X^{(1,5)}_6$ and $X^{(1,5)}_7$ reduce to the form
\beq
X_6 = \left(\ba{cccc}
v_{11} &   0    & \cdots &  0     \\ 
0      & v_{22} &        &        \\
\vdots &        & \ddots &        \\ 
0      &   0    && v_{QQ} \ea\right)\,,\qquad
X_7 = \left(\ba{cc}
w_{11}     & |w_{1\t a}|  \\ 
           &              \\ 
|w_{1\t a}|& \quad\Big(\,w_{\t a \t b}\,\Big)
\ea\right)\,.
\label{II-eleven}
\eeq
Here, $\t a, \t b = 2,\cdots, Q$ where $Q$ is either $Q_1$ or $Q_5$.

Out of the surviving $U(1)\times U(1)$ gauge group, only a diagonal
$U(1)$ subgroup has a non-trivial action on the fields $\chi_i$ and,
therefore, can be used to gauge away a single phase. We use this to
gauge away the phase of the component $(\chi_1)_{1'1}$, leaving us
with only three degrees of freedom in $\chi_{i1'1}$. The other
diagonal $U(1)$ subgroup does not transform the hypermultiplets and is
not broken. This is not unexpected since the scalar component of the
vector multiplet associated with this $U(1)$ corresponds to the center
of mass position of the combined D-brane system in the physical
4-dimensional space and hence should remain massless. Thus, after
gauge fixing, the adjoint hypermultiplets $N^{(1)}_i$ and $N^{(5)}_i$
contain $3Q_1^2 +1$ and $3Q_5^2 +1$ degrees of freedom, respectively,
and the bi-fundamental hypermultiplet $\chi_i$ contains $4Q_1Q_5-1$
degrees of freedom.

Now we consider the constraints imposed on these fields by the
vanishing of the D-terms. The trace equations (\ref{II-twelve}) can be
used to determine $|(\chi_i)_{1'1}|$ and the phase of $(\chi_2)_{1'1}$ in
terms of other $\chi$'s. This, along with gauge fixing, completely
determines $\chi_{i1'1}$, leaving us with $4Q_1Q_5 -4$ degrees
of freedom in the $\chi$ hypermultiplet.

Removing the trace parts from the D-term constraints, we are left with
$3Q_1^2+3Q_5^2-6$ equations that can be used to determine that many
components of $N^{(1,5)}_i$ in terms of the $\chi_i$. This leaves $8$
components (two adjoint hypermultiplets) undetermined. These are easy
to identify: The D-term conditions involve only the commutators of
$X^{(1,5)}_m$ (or of $N^{(1,5)}_i$) and therefore do not constrain
their traces, $x^{(1,5)}_m={\rm Tr}\, X^{(1,5)}_m$, which are the
undetermined hypermultiplets. The fields $x^{(1)}_m$ correspond to the
center of mass location of the D1-branes inside the D5-branes while
$x^{(5)}_m$ are Abelian Wilson lines on $T^4$ that have a similar
interpretation in a dual theory in which the 1-branes and the 5-branes
are interchanged. Clearly, these positions are arbitrary which
justifies the existence of the two associated massless
hypermultiplets. Therefore, with these conventions, the space of
independent field components is parametrized by the $Q_1Q_5 -1$
hypermultiplets $\chi_{ia'a}$ (excluding $a'=a=1$) and two adjoint
hypermultiplets $x^{(1)}_m$ and $x^{(5)}_m$ 
\footnote{This parametrization of the independent degrees of freedom
in terms of $\chi_{ia'a}$ (excluding $a'=a=1$) is not globally valid
and breaks down when the matrices $N_i$ commute with each
other. However, in such a situation one can find a different
parametrization. One can choose the diagonal elements of $N^{(1,5)}_i$
as independent degrees of freedom \cite{Wit2} and instead regard
$\chi_{i\t a'1}$ and $\chi_{i1'\t a}$ as being fixed by the D-term
constraints. This again leads to spaces of the form ${\cal M}_0$ in
(\ref{M-0}) or ${\cal M}$ in (\ref{II-thirteen}) below.}. 
Since $X^{(1,5)}_m$ are compact variables, it is natural to also take
$\chi_i$ to be compact so that the moduli space can be consistently
parametrized in terms of $\chi_i$. Explicitly, this space can be
written as  
\beq
{\cal M}_0= (T^4)^{\t Q_1}\times (T^4)^{\t Q_5}\times 
(T^4)^{\t Q_1\t Q_5}\times (T^4)^2\,,
\label{M-0}
\eeq
where, $\t Q_{1,5}= Q_{1,5}-1$.

The parametrization of the ($4Q_1Q_5-4$)-dimensional subspace of
fields in terms of the hypermultiplets $\chi_{i\t a'1}$, $\chi_{i1'\t
a}$ and $\chi_{i\t a'\t a}$ still has a redundancy. This is because
our gauge fixing has not properly taken care of a subgroup of the Weyl
group of the gauge group. The Weyl group of $U(Q)$ is the symmetric
group $S(Q)$, the elements of which act as permutations on the
fundamental representation index $a=1,2,\cdots,Q$ of $U(Q)$. Let us
denote by $S(Q-1)$ the subgroup that leaves $a=1$ unchanged and acts
as permutations only on elements labelled by $\t a=2,3,\cdots,Q$. We
are interested in the subgroup $S(Q_1-1)\times S(Q_5-1)$ of Weyl
reflections in $U(Q_1)\times U(Q_5)$. The elements of this subgroup
act on the rectangular matrices $\chi_{i\t a'\t a}$ by permuting their
rows and columns amongst themselves. They also act as permutations on
$\chi_{i\t a'1}$ and $\chi_{i1'\t a}$. Denoting the elements of
$S(Q_{1,5}-1)$ by $S_{1,5}$, this action can be written as
\beq
S_1\,\left(\ba{cc}
\chi_{1'1}     & \chi_{1' \t a}  \\ 
\chi_{\t a'1}   & \chi_{\t a'\t a} 
\ea\right)\, S^\d_5\,=
\left(\ba{cc}
\chi_{1'1}           & \chi_{1' p_5(\t a)}  \\ 
\chi_{p_1(\t a')1}   & \chi_{p_1(\t a')p_5(\t a)} \ea\right)\,,
\label{II-twelve'}
\eeq
where, $p_{1,5}(a)$ implement the corresponding permutations on the
index $a$. Note that the form (\ref{II-eleven}) of the matrices 
$X^{(1,5)}_m$, as dictated by our choice of gauge fixing, is
manifestly invariant under this subgroup of Weyl reflections.
As a result, the points on the space of $\chi$'s related by these 
transformations are gauge equivalent and, therefore, have to be 
identified. Taking this into account, the moduli space has the
structure  
\beq
{\cal M}= \frac{(T^4)^{\t Q_1}}{S(\t Q_1)}\times \frac{(T^4)^{\t Q_5}}
{S(\t Q_5)}\times\frac{(T^4)^{\t Q_1\t Q_5}}{S(\t Q_1)\times S(\t Q_5)}
\times (T^4)^2\,.
\label{II-thirteen}
\eeq

The dependence of $N^{(1,5)}$ on $\chi$ induces a metric on the moduli
space ${\cal M}$ (or the space ${\cal M}_0$) through the kinetic
energy terms in (\ref{one-IV}). For convenience, let us parametrize
the complex fields $\chi_i$ in terms of real fields $Y^m, (m=6,7,8,9)$
given by   
\beq
\chi_1 = Y^9+i Y^8\,,\qquad \chi_2 = Y^7+i Y^6\,,
\label{Ychi}
\eeq 
and also parametrize $N^{(1,5)}_i$ in terms of $X^{(1,5)}_m$ as in
(\ref{XN}). Then, for example, the kinetic energy term for $N^{(5)}_i$
can be written as 
\beq
\del_\al X^{(5)}_{lab}\del^\al X^{(5)}_{lab}=
G_{(5)}^{(mc'd)(ne'f)}(Y) \del_\al Y_{mc'd}\del^\al Y_{ne'f}\,,
\label{III-ten}
\eeq
with,
\beq
G_{(5)}^{(mc'd)(ne'f)}(Y)=\frac{\del X^{(5)}_{lab}}{\del Y_{mc'd}}
\frac{\del X^{(5)}_{lab}}{\del Y_{ne'f}}\,,
\label{III-eleven}
\eeq
There are similar contributions from $N^{(1)}$ and $\chi_{i1'1}$. 
Putting all this together, the metric on the $4(Q_1Q_5-1)$-dimensional
subspace of the target space spanned by $\chi_i$ takes the form 
\beq
G_{(mc'd)(ne'f)} 
=\delta_{mn}\delta_{c'e'}\delta_{df}
+\frac{k_{11}}{k_{15}}G^{(1)}_{(mc'd)(ne'f)} 
+\frac{k_{55}}{k_{15}}G^{(5)}_{(mc'd)(ne'f)} 
+G^{(\chi)}_{(mc'd)(ne'f)} 
\label{metric}
\eeq
However, note that the vanishing of the D-terms (\ref{B-III}) and
(\ref{B-IV}) imply that $G^{(1)}$ is proportional to $k_{15}/k_{11}$
and $G^{(5)}$ is proportional to $k_{15}/k_{55}$. Therefore, the
metric $G$ is independent of the couplings. Restricted to the
non-trivial $4(Q_1Q_5-1)$-dimensional part of the moduli space, the
action $S_{ke}$ in (\ref{one-IV}) becomes
\beq
S'_{ke}=k_{15}\int d^2\s G^{(mc'd)(ne'f)}(Y)\del_\al Y_{mc'd}
\del^\al Y_{ne'f}\,.
\label{Ske}
\eeq
where, $Y_{ma'a}$ does not include $Y_{m1'1}$.  

While we do not calculate the metric $G$, we can make some important
general observations: (i) In 1+1 dimensions, a hypermultiplet
decomposes into a pair of chiral multiplets of N=2 supersymmetry in
D=2, while a vector multiplet decomposes into a chiral and a twisted
chiral multiplet \cite{GHR}. A non-linear $\s$-model in two
dimensions, that admits a superfield representation, can have an
antisymmetric tensor filed coupling (the analogue of $B_{mn}$ in
string theory) only if it contains twisted chiral
multiplets. Therefore, since ${\cal M}$ is a hypermultiplet moduli
space, an antisymmetric field and hence a torsion is not induced on
it.  (ii) As a consequence of $N=4$ supersymmetry and the absence of
torsion, the metric on ${\cal M}$ is hyperkahler. (iii) Up to now our
analysis pertained to the weak coupling regime $gQ_{1,5}<1$ of the
gauge theory. However, the non-renormalization of the hypermultiplet
moduli space in gauge theory implies that the classical moduli space
discussed in the this section is not affected as we increase the gauge
coupling (see for example \cite{SEIBERG} for a statement of this). 
This means that the analysis is valid even when $gQ_{1,5}>1$.

To summarize this section, we have gauge fixed and isolated the
independent degrees of freedom which satisfy the D-flatness
conditions. These are components $\chi_{ia'a} (\sigma,t)$ (excluding
$a'=a=1$) of the (1,5) hypermultiplets along with $x_m^(1,5)$. The
remaining (1,1) and (5,5) hypermultiplets (on which we fixed the
unitary gauge) are determined in terms of these independent
components. The slowly varying independent moduli are described by a
non-linear sigma-model on a $(4Q_1Q_5+4)$-dimensional target space
whose validity is ensured even when $gQ_{1,5}>1$. In the next section,
we describe very low-energy excitations of these moduli fields.

\section{Discrete Gauge Symmetry and Low Energy Degrees of Freedom}

In this section we describe the mechanism by which very low-energy
excitations of the moduli fields, to be identified as the black hole
degrees of freedom, arise. One approach to the study of this problem
would be to consider the non-linear sigma-model on the orbifold space
${\cal M}$ as given in (\ref{II-thirteen}), with the induced metric on
it. The approach we follow is to consider the sigma-model on ${\cal
M}_0$, as given by (\ref{M-0}), and regard the residual Weyl symmetry
$S(Q_1-1)\times S(Q_5-1)$ as a discrete gauge group to be implemented
as a Gauss law constraint on the physical states. In the presence of
this discrete gauge group, the theory develops sectors with
twisted-periodic boundary conditions, leading to excitations with
fractional momenta \cite{DVV2,DVV1,DVV3,MOTL}. The effective theory
for these modes is argued to be a $c=6$ superconformal field theory
which emerges in the infra-red limit.

Consider $\chi_{ia'a} (\sigma,t)$ (excluding $a'=a=1$) as taking
values in $T^{4(Q_1Q_5-1)}$ with an $S(Q_1-1)\times S(Q_5-1)$ action
on it as given by(\ref{II-twelve}). Since configurations related by
this residual gauge group are to be identified, the compactness of
$\s$ allows us to impose twisted periodicity conditions on the
components of the matrix $\chi$, so that 
\beq
\chi_i(\sigma+ 2\pi R) = S_1 \chi_i(\sigma) S^\dagger_5\,.
\label{III-one}
\eeq
$S_1$ acts as a permutation on the row index $\t a'=2,\cdots, Q_1$ and
we denote its action by $p_1(\t a')$. Similarly, we denote a
permutation of the column index $\t a=2,\cdots,Q_5$, under the action
of $S_5$, by $p_5(\t a)$. Then, in terms of components, the twisted
boundary condition above breaks up into
\bea
\chi_{i\t a'1}(\sigma+ 2\pi R) &=& \chi_{ip_1(\t a') 1}(\sigma)\,, 
\label{III-twoI}\\
\chi_{i1'\t a}(\sigma+ 2\pi R) &=& \chi_{i1'p_5(\t a)}(\sigma)\,,
\label{III-twoII}\\
\chi_{i\t a'\t a}(\sigma+ 2\pi R) &=& \chi_{ip_1(\t a')p_5(\t
a)}(\sigma)\,.
\label{III-twoIII}
\eea
Note that $\chi_{i1'1}$, which is not an independent parameter, is
invariant.  The theory develops different sectors depending on the
elements $S_{1,5}$ chosen to implement the twisted boundary
conditions.  In each sector, defined by a choice of $S_1$ and $S_5$,
all elements of $\chi_i$ that are related to each other by the twisted
boundary condition, can be sewn into a single function defined on a
circle of larger radius, as explained below. These sewn functions are
the natural variables corresponding to very low-energy excitations
of the D-brnane system.

The elements of the symmetric group $S(Q)$ fall into conjugacy classes
that are characterized by the {\it order} and {\it number} of cyclic
subgroups that an element can generate. Suppose that one of the cyclic
subgroups generated by $S_1\in S(Q_1-1)$ is of order $l_1$. This
cyclic group will permute $l_1$ rows of the matrices $\chi_i$ amongst
themselves (keeping the first row invariant). Let us consider one of
these rows labelled by the fundamental representation index $\t
a'_0$. After going once around the compact direction, this row index
is transformed into $\t a'_1= p_1(\t a'_0)$, and after going $n$ times
around this $S^1$, we get $\t a'_n= p_1 (\t a'_{n-1})= p_1^n(\t
a'_0)$. Clearly, $\t a'_{l_1}=p_1^{l_1}(\t a'_0)=a'_0$. A similar
discussion applies to a column index $\t a_0$ transforming under a
permutation group of order $l_5$ which is one of the cyclic groups
generated by $S_5$. To describe the sewing procedure for the elements
of $\chi$, let us first consider the simpler case involving the
elements $\chi_{i\t a' 1}$ that transform only under $S_1$, and are
invariant under $S_5$. Consider a specific matrix element labelled by
$(\t a'_0,1)$ that transforms under a cyclic subgroup of order $l_1$.
As we go around the circle $l_1$ times, the twisted boundary condition
(\ref{III-twoI}) forces the function $\chi_{i\t a'_0 1}$ to go
through the $l_1$ functions $\chi_{i\t a'_n 1}$, for $n=1,\cdots,l_1$,
before coming back to itself. To sew these $l_1$ functions, each
defined on a circle of radius $R$, into one single function
$\t\chi_{i(l_1)}$ defined on a circle of radius $Rl_1$, let $\t\sigma$
be a parameter along the larger circle ($0\le\t\sigma\le 2\pi
Rl_1$). Then, $\t\chi_{i(l_1)}(\t\sigma)$ can be defined such that 
\beq
\t\chi_{i(l_1)}(2\pi R n\le \t\sigma \le 2\pi R (n+1)) = 
\chi_{ip_1^n(\t a'_0) 1}(\sigma) 
\equiv  \chi_{i\t a'_n 1}(\sigma)\,,
\label{III-three}
\eeq
where $\t\sigma=2\pi R n+\sigma$.

The kinetic energy terms for the above components of $\chi_i$, that
have been sewn into a single function, can be easily expressed in terms
of the sewn field. Using (\ref{III-three}), it is easy to see that
(suppressing the $SU(2)_R$ index $i$) 
\bea
\int_0^{2\pi Rl_1} d\t\sigma \del_\al\t\chi_{(l_1)}(\t\sigma)
\del^\al \t\chi^*_{(l_1)}(\t\sigma)
&\equiv& 
\sum_{n=0}^{n=l_1-1}\int_{2\pi Rn}^{2\pi R(n+1)}d\t\sigma
\del_\al\t\chi_{(l_1)}(\t\sigma)\del^\al\t\chi^*_{(l_1)}(\t\sigma)
\nonumber\\
&=&\sum_{n=0}^{n=l_1-1}\int_0^{2\pi R}d\sigma
\del_\al\chi_{a'_n 1}(\sigma) \del^\al\chi^*_{a'_n 1}(\sigma)\,.
\label{III-four}
\eea
In general, $S_1$ can generate many cyclic groups of various orders
$l_1$ such that $\sum_{\{l_1\}}l_1=Q_1-1$. Then, all the $Q_1-1$
elements $\chi_{\t a'1}$ can be sewn into a smaller number of
functions $\t\chi_{(l_1)}$, each living on circle of larger radius
$Rl_1$. Hence, the part of the kinetic energy term involving these
elements can be written entirely in terms of the sewn fields as
\beq 
\sum_{\t a'=2}^{Q_1}\int dt\int_0^{2\pi R}d\sigma 
\del_\al \chi_{i\t a'1}\del^\al\chi^{*}_{i\t a'1}
=\sum_{\{l_1\}}\int dt\int_0^{2\pi Rl_1}d\t\sigma \del_\al
\t\chi_{i(l_1)}\del^\al\t\chi^{*}_{i(l_1)}\,. 
\label{III-five}
\eeq
The most important outcome of the sewing procedure is that, while the
compact dimension has a physical radius $R$, the momenta of the sewn
fields $\t\chi_{(l_1)}$ are quantized in units of $1/Rl_1$. 
Furthermore, the number of independent fields reduces from $4(Q_1-1)$
for $\chi_{i\t a'1}$ to the number of sewn fields $\t\chi_{i(l)}$. 
This number is given by the number of cyclic groups generated by 
$S_1$ and could be much smaller than $4(Q_1-1)$. Obviously, the above
procedure carries over, without any change, to the elements
$\chi_{1'\t a}$ that are permuted by $S^\d_5$, but are invariant under
$S_1$. 

Having demonstrated the sewing procedure in a simpler situation, let
us now consider a matrix element labelled by $(\t a'_0,\t a_0)$, where
the first index transforms under a cyclic group of order $l_1$ and the
second, under one of order $l_5$. Due to the twisted boundary
condition (\ref{III-twoIII}), after going around the circle $n$ times,
this element is transformed into one labelled by $(\t a'_n,\t
a_n)=(p_1^n(\t a'_0), p_5^n(\t a_0))$. Clearly, the element $(\t
a'_0,\t a_0)$ comes back to itself after going around the circle $l$
times, where $l$ is the {\it least common multiple} of $l_1$ and
$l_5$. This means that the function $\chi_{i\t a'_0 \t a_0}$ goes
through $l$ functions $\chi_{i\t a'_n \t a_n}$, for $n=1,\cdots,l$,
before coming back to itself. These $l$ functions, each defined (but
not periodic) on a circle of radius $R$, can be sewn into one single
function $\t\chi_{i(l)}$ defined on a circle of radius $Rl$ as before:
Let $\t\sigma$ be a parameter along the larger circle
($0\le\t\sigma\le 2\pi Rl$). Then, $\t\chi_{i(l)}(\t\sigma)$ can be
defined such that 
\beq
\t\chi_{i(l)}(2\pi R n\le \t\sigma \le 2\pi R (n+1))= 
\chi_{ip_1^n(\t a'_0) p_5^n(\t a_0)}(\sigma) 
\equiv
\chi_{i\t a'_n \t a_n}(\sigma)\,,
\label{III-six}
\eeq
where $\t\sigma=2\pi R n+\sigma$. Using this, the kinetic energy terms
for the sewn components of $\chi_{\t a'\t a}$ can be written in terms
of the single $\t\chi_{(l)}$ as (again, suppressing the $SU(2)_R$
index)   
\bea
\int dt\int_0^{2\pi Rl} d\t\sigma \del_\al\t\chi_{(l)}(\t\sigma)
\del^\al \t\chi^*_{(l)}(\t\sigma)
&\equiv&
\sum_{n=0}^{n=l-1}\int dt \int_{2\pi Rn}^{2\pi R(n+1)}d\t\sigma
\del_\al\t\chi_{(l)}(\t\sigma)\del^\al\t\chi^*_{(l)}(\t\sigma)
\nonumber\\
&=&\sum_{n=0}^{n=l-1}\int dt\int_0^{2\pi R}d\sigma
\del_\al\chi_{a'_n a_n}(\sigma) \del^\al\chi^*_{a'_n a_n}(\sigma)\,.
\label{III-seven}
\eea

Depending on the structure of $S_1$ and $S_5$, all the
$(Q_1-1)(Q_5-1)$ complex elements $\chi_{i\t a'\t a}$, defined on a 
circle of radius $R$, can be sewn into a smaller number of functions
$\t\chi_{i(l)}$ defined on circles of radii $Rl$, such that
$\sum_{\{l\}}l= (Q_1-1)(Q_5-1)$. The kinetic energy terms for these
components can then be written as
\beq 
\sum_{\t a'=2}^{Q_1}\sum_{\t a=2}^{Q_5}\int dt\int_0^{2\pi R}
d\sigma \del_\al \chi_{i\t a'\t a}\del^\al\chi^{*}_{i\t a'\t a} 
=\sum_{\{l\}}\int dt\int_0^{2\pi Rl}d\t\sigma 
\del_\al\t\chi_{i(l)}\del^\al\t\chi^{*}_{i(l)}\,. 
\label{III-eight}
\eeq
The field $\t\chi_{(l)}$ has momentum quantized in units of $1/Rl$.
Of course, if we are interested in long wavelength effects, then only
sectors with the largest $l$ will contribute (these sectors are also
more favourable entropically). 
 
A sewn field $\t\chi$ with the lowest quantum of momentum can be
obtained if $\t Q_1=Q_1-1$ and $\t Q_5=Q_5-1$ are coprime and
$S_{1,5}$ in (\ref{III-one}) are chosen such that they generate cyclic
groups of order $\t Q_{1,5}$. This can be easily achieved by choosing
$S_{1,5}$ to cyclically permute the indices $(\t a',\t a)$: $(p_1(\t
a'),p_5(\t a))=(\t a'+1,\t a+1)$. In this case, using (\ref{III-six}),
all the $4\t Q_1 \t Q_5$ real components of $\chi_{i\t a' \t a}$ can
be sewn into a $4$ real functions $\t\chi_i$ that are periodic on a
circle of radius $R\t Q_1\t Q_5$, and thus have their momenta
quantized in units of $1/R\t Q_1\t Q_5$. The kinetic energy term for
these components of $\chi$ takes a very simple form: 
\beq
\sum_{\t a'=2}^{Q_1}\sum_{\t a=2}^{Q_5}\int dt\int_0^{2\pi R}d\sigma 
\del_\al \chi_{i\t a'\t a}\del^\al\chi^{*}_{i\t a'\t a}
=\int dt\int_0^{2\pi R\t Q_1\t Q_5}d\t\sigma 
\del_\al\t\chi_i\del^\al\t\chi^{*}_i\,.
\label{III-nine}
\eeq

With this twisted boundary condition, the elements $\chi_{\t a'1}$
($\chi_{1'\t a}$) can be sewn into a function $\t\chi_{(\t Q_1)}$
($\t\chi_{(\t Q_5)}$) with momentum quantized in units of $1/R\t Q_1$
($1/R\t Q_5$). The remaining two moduli fields $x^{(1)}_m$ and
$x^{(5)}_m$ are periodic on the circle of radius $R$. Ignoring the
contribution of the fields $N^{(1,5)}$ for the time being, it is
reasonable to expect that for large $Q_1$ and $Q_5$ only the sewn
field appearing in (\ref{III-nine}), with 4 real components, would be
relevant to the physics of the black hole at low energies. In case $\t
Q_1$ and $\t Q_5$ are not coprime, one can choose $S_1$ and $S_5$ such
that they contain cyclic subgroups of maximum possible order $l_1$ and
$l_5$ which are coprime. The sewn field $\t\chi_{(l_1l_5)}$ coming
from this sector will then have the lowest momentum quantum and will
dominate the low-energy physics. For sectors corresponding to other
conjugacy classes of the Weyl group, one can construct tunnelling
configurations in the gauge theory that could induce transitions
between these sectors. As a result, smallest cycles may go over to
longer cycles for reasons of phase space, because the entropy
associated with long cycle is not less than the sum of entropies of 
smaller cycles. 
 
So far, we have only considered the moduli fields $\chi_{ia'a}$
(excluding $\chi_{i1'1}$) and their kinetic energy terms. The full
kinetic energy action (\ref{one-IV}) also contains the fields
$N^{(1,5)}_i$ and $\chi_{i1'1}$ which give rise to a metric on the
moduli space through their dependence on the $\chi$-moduli as in
(\ref{Ske}).  On imposing twisted boundary conditions, the presence of
$G_{(mc'd)(ne'f)}$ leads to {\it non-local} interactions for the sewn
fields $\t\chi_{(l)}$ (or equivalently, for $\t Y_{(l)}$) in the
theory with a $4$-dimensional target space defined on the circle of
radius $Rl$. This is not difficult to see: The index $(a'a)$ of an
element of $\chi_i$ determines a small interval on the larger circle
on which $\chi_{ia'a}=\t\chi_{i(l)}$. Therefore, $\chi_{ic'd}$ and 
$\chi_{ie'f}$ (provided they are in the same sector of the twisted
boundary condition) may correspond to the values of some
$\t\chi_{i(l)}$ at different points on the circle of radius $Rl$.  If
$\chi_{ic'd}$ and $\chi_{ie'f}$ are not in the same sector, then $G$
may introduce couplings between different sectors which may have their
fractional momenta quantized in different units. The metric will also
introduce couplings between the components $\chi_{i\t
a'1}$,$\chi_{i1'\t a}$ and $\chi_{i\t a'\t a}$.

Thus, the full theory in terms of the sewn variables is a very
complicated, non-local sigma-model. However, we are only interested in
the lowest energy excitations on the moduli space. To isolate this
sector, we have to integrate out all the higher momentum sectors,
retaining the lowest one, say $\t\chi_{i(\t Q_1\t Q_5)}$ (where, for
simplicity, we assume that $\t Q_1$ and $\t Q_5$ are
coprime). Although this cannot be done explicitly, this procedure
leads to an effective theory for the renormalized form of
$\t\chi_{i(\t Q_1\t Q_5)}$ which we denote by $\t\chi_i$, and which
lives on a circle of radius $R\t Q_1\t Q_5$. Essentially, $\t\chi_i$ is
an order parameter in terms of which the infra-red physics can be
described. While we cannot obtain the effective low-energy theory for
$\t\chi_i$ by explicitly integrating out the higher momentum sectors,
we make the assumption that this theory is local to lowest order in
energy. This implies that we are dealing with a non-linear sigma-model
on a compact $4$-dimensional manifold,
\beq
k_{15}\int dt\int_0^{2\pi R\t Q_1\t Q_5}  d\t\sigma
G_{mn}(\t Y)\del_\al\t Y^{m}\del^\al\t Y^{n}\,,
\label{GYY}
\eeq
where, $\t Y^m, (m=6,7,8,9)$ are the real components of $\t\chi_i$
given by $\t\chi_1 = \t Y^9+i\t Y^8$ and $\t\chi_2 = \t Y^7+i\t Y^6$.
$N=4$ supersymmetry, which is expected to survive in the
infra-red limit, implies that this manifold is hyperkahler \cite{Luis}
and therefore, it is either $T^4$ or a $K3$ surface. This great
simplification, over the original complicated non-local theory, is a
consequence of N=4 supersymmetry and compactness of the moduli
space ${\cal M}_0$, together with our assumption of the locality of
the infra-red theory. To make a choice between $T^4$ and $K3$, let us
consider $K3$ in the orbifold limit $T^4/Z_2$. The modding by $Z_2$
identifies $\t Y^m$ with $-\t Y^m$, whereas, such an identification is
not required in the original gauge theory, nor it is imposed by the
D-term constraints or the infra-red limit. While, strictly speaking,
this argument only rules out $T^4/Z_2$, it may also be an indication
of the existence of more general obstructions to $K3$. There is also
an important {\it a posteriory} argument in favour of $T^4$: Locally,
the metric on $T^4$ has an $SO(4)$ invariance which is crucial in
determining the couplings of the minimal scalars to the low-energy
excitations of the D-brane system, leading to the correct calculation
of Hawking emission and absorption rates. Therefore, it is reasonable
to regard (\ref{GYY}) as a sigma model on $T^4$. The emergence of the
$SO(4)$ invariance is then a consequence of the infra-red limit.

Clearly, the compactness of the moduli space is responsible for
restricting our choice of the hyperkahler metrics to $T^4$ and $K3$.
If the moduli space ${\cal M}_0$ in (\ref{M-0}) is not compact (as
would be the case for a generic gauge theory), that is, if the factors
of $T^4$ in (\ref{M-0}) are replaced by $R^4$, then the dynamics of
the order parameter is governed by a sigma-model on a $4$-dimensional 
non-compact hyperkahler, and hence Ricci flat, manifold. In order to
deduce further implications of the vanishing of the Ricci tensor, we
need to specify the boundary conditions on the metric on this space. 
If we make the simplest assumption that, for large $\t Y^m$, this
space has the structure of the 4-dimensional Euclidean space, then a
theorem by Witten \cite{Wit3} states that this space is a flat
4-dimensional Euclidean space. Furthermore, an argument similar to the
one ruling out $T^4/Z_2$ may be used to rule out spaces with
non-trivial topology at infinity. In spite of this, there are many
more choices than in the compact case.

To summarize, in this section we have shown that the residual discrete
gauge invariance of the theory leads to fractional excitations of the
moduli fields. We then identified an order parameter $\t\chi_i$ (or
$\t Y^m$), periodic on a circle of radius $R\t Q_1\t Q_5$, in terms of
which the low-energy dynamics of the system in the infra-red limit can
be described. The corresponding effective theory, including the
fermions, is a $c=6$ superconformal field theory on $T^4$ with
extended $N=4$ supersymmetry. The $SU(2)\times SU(2)$ R-symmetry 
of this superconformal algebra acts on the fermionic partners of $\t
Y_m$ and is identified as the $SO(4)$ group of rotations in the
physical 4-dimensional space transverse to $S^1\times T^4$ \cite{ANG}.
Therefore, fermionic excitations in the superconformal field theory
carry angular momentum corresponding to rotations in the physical
space transverse to the D-brane system. In the next section, we will
focus on the relation between this effective theory and the
5-dimensional black hole.  

\section{The Black Hole, Coupling to Bulk Fields and Hawking
Radiation} 

We have shown that the D1,D5-brane system has low-energy degrees
of freedom with fractional momenta, which in the infra-red limit, are
effectively described by a $c=6$ superconformal field theory. Writing
only the the bosonic terms,
\beq 
S_{eff}= T_{eff}\int dt\int_0^{2\pi R\t Q_1\t Q_5} d\t\sigma
\del_\al\t Y^{m}\del^\al\t Y^{m}\,,
\label{Seff}
\eeq
where, up to a normalization, $T_{eff}$ is given by
\beq
T_{eff}= k_{15} = \frac{1}{\al'^2g}\sqrt{\frac{V_4}{Q_1Q_5}}\,.
\label{Teff}
\eeq 
As discussed in the introduction, a theory of this form, usually
called the ``effective string picture'', has been used to model the
5-dimensional black hole with rather good success. However, in our
case, the $T^4$ on which the sigma-model is defined does not have a
space-time interpretation and, in particular, is not the $T^4$ on
which the D5-brane is compactified. Thus, strictly speaking, our $T^4$
is not the target space of some effective string, to be regarded as
some kind of D-string, though in some cases this picture can prove
very useful.  The obtaining of the model (\ref{Seff}) from the
D-brane gauge theory not only explains its success but, more
importantly, sheds light on some of its less understood features as
will be discussed below. Moreover, this approach puts the calculation
of the Hawking radiation and the resolution of the information paradox
for these black holes on a more solid foundation.  It may also lead to
an understanding of higher angular momentum \cite{Gu,MATHUR}
processes, though we will not dwell on this further.

Our derivation relates the coupling $T_{eff}$, often called the
``effective string tension'', to the (1,5) hypermultiplet coupling
$k_{15}$ (\ref{Teff}), which was determined in section 2. As a check,
this can be compared with the behaviour of $T_{eff}$ that one expects
in order to get agreement between various cross sections calculated in
the SCFT framework and the corresponding ones obtained by
semi-classical black hole calculations. $T_{eff}$ does not appear in
the cross sections for minimally coupled scalars calculated in the
$c=6$ superconformal field theory, though the ones for fixed scalars
do depend on it \cite{CGKS,KK}. It is also argued to appear in the
cross sections for higher angular momentum modes of the minimal
scalars, to the extent that their structure can be surmised within
this SCFT approach \cite{MATHUR}. In fact, in \cite{MATHUR}, the
dependence of $T_{eff}$ on various parameters of the theory, notably
its dependence on $V_4$, was suggested so that the SCFT has the
ability to reproduce the cross sections for higher angular momentum
modes in agreement with semi-classical black hole calculations. The
form of $T_{eff}$ we have obtained is in agreement with these
calculations, as well as with the fixed scalar calculations, where
they can be performed.

In the SCFT (\ref{Seff}), the extremal 5-dimensional black hole can be
identified with left moving states $|N_L>$ at certain level $N_L$ in
the standard way \cite{SEN,StVa,CaMa,MaSu}. From the expansion
\beq
\t Y^m (\t\s,t) = \sum_{p} \al^m_p e^{i (p/R\t Q_1\t Q_5)(\t\s +t)}
+ (\mbox{right moving part})\,,
\label{mod}
\eeq
it is clear that a state at level $N_L$ is associated with total
left-moving momentum $N_L/R\t Q_1\t Q_5$. Since the black hole charge
$N$ is related to the left-moving momentum $N/R$ along $x^5$, the
states corresponding to it in the CFT should be at level $N_L = N\t
Q_1\t Q_5$. These states preserve the same amount of supersymmetry as
the classical black hole solution. The entropy is then related to the
degeneracy of states at this level and for large $N_L$ is
$S=2\pi\sqrt{N_Lc/6}=2\pi\sqrt{N \t Q_1\t Q_5}$. For large $Q_{1,5}$
this agrees with the Hawking-Bekenstein entropy $2\pi\sqrt{NQ_1Q_5}$
for this black hole. It is important to recall that, as argued in
section 3, the moduli space analysis there could be extrapolated from
the perturbative gauge theory regime of $gQ_{1,5}<1$ to the large
black hole regime of $gQ_{1,5}>1$. The sigma-model (\ref{Seff}) is
therefore valid in the large black hole regime (otherwise,
$T_{eff}\rightarrow \infty$ as $gQ_{1,5}\rightarrow 0$ and it becomes
almost impossible to excite the black hole away from extremality).

The black hole can be perturbed away from extremality, keeping its
macroscopic charges unchanged, by adding small and equal amounts of
right and left moving fractional momenta along $x^5$ \cite{CaMa,MaSu}
(we do not consider going off extremality by adding anti-branes). In
the conformal field theory, the corresponding states are $|N_L+\delta
N_R>\otimes |\delta N_R>$. These states are no longer BPS saturated
and are not protected by supersymmetry. However, based on the
non-renormalization of the hypermultiplet moduli space, we expect that
the basic SCFT description is valid even at strong coupling and
therefore, the correspondence between these non-BPS states and
non-extremal black holes holds even in the macroscopic black hole
limit. This expectation is supported by the accuracy with
which Hawking radiation calculations can be performed. In general, in
the vicinity of the extremal black hole states $|N_L>\otimes |0>$ with
$N_L=N\t Q_1\t Q_5$, the CFT could also contain states like
$|N_L>\otimes |\delta N_R>$ that do not have an interpretation in
terms of the original black hole. If the SCFT is to provide a faithful
description of the extremal black hole and its near extremal
deformations, then it should automatically prevent such states from
appearing. This is achieved by a level matching condition that, in our
approach, is inbuilt in the conformal field theory: The residual
discrete gauge transformations (\ref{II-twelve}) of the super
Yang-Mills theory correspond to discrete translations along $\t\s$ in
(\ref{Seff}). Imposing invariance under these translations as a Gauss
law constraint implies that the allowed physical states are the ones
for which the momentum operator along $\t\s$ has integer eigenvalues,
\beq
\frac{L_0-\b L_0}{\t Q_1\t Q_5} = n \,\,\mbox{(integer)}\,.
\label{lm}
\eeq
For the extremal black hole $n=N$, and any change in $n$ corresponds
to going away from extremality by a large amount. Thus the only
allowed states in the vicinity of the extremal state are the ones for
which $\Delta (L_0-\b L_0)=0$. This condition eliminates the conformal
field theory states that do not have a counter part on the black hole
side.

As far as the relation to black hole physics is concerned, the most
important issue is the coupling of this effective $c=6$ SCFT to the
bulk fields. These couplings are responsible for the processes of
Hawking emission and absorption of the bulk fields
\cite{CaMa,DMW,DM1,MaSt,CGKS,KK,JKM,ANG}. Having identified the extremal
and near extremal black holes in terms of states in the SCFT, it is
natural to describe couplings to bulk fields by operators in this
SCFT, subject to the level matching condition. However, a first
principle derivation of the interaction between the effective SCFT and
bulk matter (closed strings states) is somewhat difficult and
therefore, as yet, there is no known mechanism of systematically
identifying these operators (though the Dirac-Born-Infeld action
approach is useful and may give some hint \cite{CGKS}). In the
remaining part of this section, we will discuss coupling to the
minimal scalars in our setup.

The simplest example of coupling to bulk fields is the coupling to the
scalar fields $h_{mn}(t,x^1,\cdots x^9), (m\neq n)$ that correspond to
the components of the 10-dimensional graviton along the
compactification $T^4$ in the directions $x^6,x^7,x^8,x^9$. Since we
are interested in low-energy processes, we restrict $h_{mn}$ to
its zero modes along $T^4$. Furthermore, the momentum of $h_{mn}$
along $x^5$ is quantized in units of $1/R$ as opposed to $1/R\t Q_1\t
Q_5$ for the modes of the effective SCFT. Therefore, at low energies,
only the zero momentum modes of $h_{mn}$ along $x^5$ couple to the
SCFT. Since, the SCFT lives at the origin of uncompactified space, we
restrict the interaction to take place at $x^1,\cdots,x^4=0$. Hence,
we only have to consider $h_{mn}(t,x^\mu=0)$.  At long wavelengths,
and to lowest order, a coupling of these minimal scalars to
(\ref{Seff}) can be written simply based on symmetry principles. We
first write down this ``phenomenological'' action and then see how it
could arise from the microscopic point of view, in the process
resolving a puzzle noted in \cite{Ha}. As already noted, in the long
wavelength limit, the $SU(2)_R$ symmetry of the super Yang-Mills
theory is enlarged to the $SO(4)$ symmetry of $S_{eff}$ in
(\ref{Seff}). The simplest form of coupling to $h_{mn}$ is basically
dictated by this $SO(4)$ and the relevant phenomenological action can
be written as
\beq
T_{eff}\int dt\int_0^{2\pi R\t Q_1\t Q_5} d\t\sigma
(\delta _{mn} + h_{mn}(t))\del_\al \t Y^{m}\del^\al\t Y^{n}\,.
\label{IV-two}
\eeq 
The origin of this coupling deserves some attention. Although, even a
simple counting argument indicates that the (1,5) hypermultiplets
$\chi_i$ are responsible for the black hole degrees of freedom
\cite{CaMa,MaTh}, the calculation in \cite{Ha} shows that these
hypermultiplets do not couple to minimal scalars at the microscopic
level. This is contrary to the expectation that these scalars should
contribute to Hawking emission and absorption by the black hole, as
suggested by semi-classical black hole calculations. We can see that
the coupling (\ref{IV-two}) in the effective low-energy theory arises
from the coupling, at the microscopic level, of gravitons to the (1,1)
and (5,5) hypermultiplets $N^{(1,5)}_i$ (or, in terms of real
components, $X^{(1,5)}_m$). Such a coupling modifies the kinetic
energy terms (\ref{one-IV}) of these hypermultiplets (now written in
terms of real components) to $(\delta^{mn}+ h^{mn})\del_\al
X^{(1,5)}_m\del^\al X^{(1,5)}_n$. This in turn, modifies equations
(\ref{III-ten})-(\ref{metric}) so that, in the infra-red theory, the
$G_{mn}$ in (\ref{GYY}) is replaced by $(\delta_{pq}+h_{pq})
G^{pq}_{mn}$. In the absence of $h_{pq}$, we argued that $G^{pq}_{mn}$
is of the form $\delta^p_m\delta^q_n$. Regarding $h_{pq}$ as a small
perturbation and hence retaining the same form for $G^{pq}_{mn}$,
leads us to the effective coupling (\ref{IV-two}).
 
Equation (\ref{IV-two}), which is valid in the regime $gQ_{1,5}>1$, is
one of our main results. From here one can compute the grey body
factors and match them with the General Relativity calculations.
These calculations, which for the scalar emission considered above, do
not depend on $T_{eff}$, have already been performed based on a model
due to Maldacena and Susskind \cite{MaSu}. The interaction lagrangian
that was used in \cite{DMW} has a form almost identical to
(\ref{IV-two}).  From the DBI action, used in \cite{DM1,CGKS}, the
same form of interaction emerges. This agreement, in spite of the
difference in the underlying pictures, is not surprising since this
coupling is more or less dictated by $SO(4)$ symmetry. 

Regarding the coupling of fixed scalars to the effective SCFT,
from \cite{KK} it is clear that this coupling involves SCFT operators 
with more than two derivatives and hence it is presently beyond the
scope of the moduli space approximation and the long wavelength limit
to which the non-renormalization theorem applies. Incorporation of the
fixed scalars will probably require a better understanding of the
model. 

One of the most important issues that any derivation of black hole
thermodynamics from string theory must face is the issue of
thermalization which, even in standard systems, is often only argued
for rather than proven. We note that the non-local interactions that
arise as a result of applying the sewing procedure to (\ref{Ske}) can
help the system thermalize relatively easily because every bit of the
``long string'' is in contact with every other bit. This would justify
the use of the canonical ensemble for the SCFT.

We end this section by discussing the range of validity of comparisons
between the D-brane and semi-classical results: Unlike Schwarzschild black
holes, the 5-dimensional black hole we have considered has a positive
specific heat $C=\del M/\del T_H\sim T_H$, where $T_H$ goes to zero
with the non-extremality parameter $r_0$ \cite{MaSt}. The temperature
fluctuations $\Delta T$ are related to the specific heat $C$ by
\cite{Pr}  
\beq
\frac{<(\Delta T)^2>}{T^2} = \frac{1}{C}\,,
\eeq
and therefore, blow up as $T_H\rightarrow 0$. Thus, as the black hole
approaches the extermal limit, the thermodynamic picture breaks down. 
For the black hole we have considered, $C\simeq V_4R^2r_0^2
\sqrt{Q_1Q_5/N}$. Therefore, the comparison with semi-classical
picture is valid only when
\beq
r_0^2 \sqrt{\frac{Q_1Q_5}{N}} >> 1\,.
\eeq
Outside this range, the thermodynamic picture breaks down while the
SCFT calculation in the D-brane framework is still valid. This problem
also manifests itself in the calculation of the entropy of the near
extremal black holes \cite{CaMa}: If the right-moving oscillator
number $\delta N_R$ is small (which it is, very close to extremality),
the density of states does not grow exponentially with $\delta N_R$
and hence does not lead to the thermodynamic entropy. This is not
surprising as the thermodynamic picture is not valid in this regime.

\section{Conclusions}

In summary, we have studied the D-brane constituent model of the near
extremal black hole of IIB string theory in 5-dimensions. We have
extracted the low-energy degrees of freedom for $gQ_{1,5}>1$ and the
corresponding effective theory which describes black hole
thermodynamics. This turns out to be an $N=4$ free superconformal
field theory with $c=6$ and an enhanced $SO(4)$ symmetry. The coupling
$T_{eff}$ is given by (\ref{Teff}). The level matching condition
(\ref{lm}) guarantees a faithful description of the near extremal
excitation of the black hole in terms of SCFT states.  We have also
discussed the phenomenological lagrangian that describes the
interaction of bulk minimal scalars with this SCFT. While the minimal
scalars do not couple to the (1,5) hypermultiplets at the microscopic
level, a coupling to the black hole degrees of freedom is induced in
the effective theory. This provides a string theoretic basis for the
Hawking decay of the excited black hole to its ground state. Now that
the derivation of the Hawking decay rate has a sound basis in string
theory we can argue that, at least in this model, there is no
information loss when the black hole radiates to its ground state by
the emission of long wavelength quanta. The black hole thermodynamics
is born of the usual statistical averaging procedure of quantum
statistical mechanics. The so called ``information loss'', which would
seem to occur in the de-excitation of a near extremal black hole in
General Relativity, is now explained as arising from the averaging
procedure of an S-matrix which knows about all the phase correlations.
This resolution is of course based on an S-matrix approach.  However
it would be of great interest to make contact with the geometry of the
black hole that arises from the D-brane model. Only then we would have
understood the Hawking-Bekenstein formula for the black hole entropy
and resolved some of the mysteries that underly Hawking's original
derivation of black hole thermodynamics \cite{HAWK}. 

There is also another approach to the study of these black holes 
\cite{Sfet-n,Mal-n,Strom-n} and it is interesting to explore the
relationship between the two associated superconformal field
theories. The idea is that by an appropriate U-duality transformation,
the D-brane black hole can be mapped into another classical solution
carrying only NS-NS charges with the structure $BTZ\times S^3\times
T^4$. This also happens to be the near-horizon geometry of the 
$D1,D5$-brane black hole. When the momentum along $x^5$ is set to
zero, this reduces to $AdS_3\times S^3\times T^4$ and is associated
with a $N=(4,4)$ WZNW model on $SL(2)\times SU(2)\times T^4$. The
black hole is then associated with states in this SCFT. Since entropy
is U-duality invariant, it can be computed in this SCFT. In this
approach one can avoid extrapolation in the string coupling as well as
the issue of coupling bulk fields to SCFT for R-R excitations. 
However, the microscopic origin of the system is not as evident as
in the D-brane black hole case. Understanding the connection between
these two approaches will certainly lead to a better understanding of
not only the black hole problem, but also some aspects of string
theory.

\vspace*{.4cm} 

\noindent{\bf\Large{Acknowledgments}}

\vspace*{.3cm} 

\noindent We would like to thank L. Alvarez-Gaume, D. Amati, I. Bakas,
S. R. Das, M. Douglas, E. Kiritsis, J. Maldacena, G. Mandal, G. Moore,
M. Moriconi, K. S. Narain, S. Randjbar-Daemi, A. Sen, G. Thompson,
G. Veneziano, E. Verlinde and E. Witten, for useful discussions and
comments during the course of this work.

\end{document}